\documentclass{article}
\usepackage{iclr2025_conference,times}

\usepackage{graphicx}
\usepackage{subfigure}
\usepackage{multirow}
\usepackage{tcolorbox}
\usepackage{listings}
\usepackage{wrapfig}
\usepackage{hyperref}
\usepackage{url}
\usepackage[utf8]{inputenc} 
\usepackage[T1]{fontenc}    
\usepackage{booktabs}       
\usepackage{amsfonts}       
\usepackage{nicefrac}       
\usepackage{microtype}      
\usepackage{xcolor}         

\title{AstroLLaVA: Towards the Unification of\\ Astronomical Data and Natural Language\\}

\newcounter{affcounter}

\newcommand{\aff}[1]{%
  \ifcsname aff@#1\endcsname%
    \csname aff@#1\endcsname%
  \else%
    \stepcounter{affcounter}%
    \expandafter\xdef\csname aff@#1\endcsname{$^{\theaffcounter}$}%
    \csname aff@#1\endcsname%
  \fi%
}

\author{\begin{tabular}{l}
Sharaf Zaman\aff{UniverseTBD}$^{,*}$ \quad Michael J. Smith\aff{UniverseTBD}$^{,\dagger}$ \quad Pranav Khetarpal\aff{UniverseTBD}$^,$\aff{Indian Institute of Technology, Delhi} \\[4pt] 
Rishabh Chakrabarty\aff{UniverseTBD}$^,$\aff{ Intelligent Internet Inc.} \quad Michele Ginolfi\aff{UniverseTBD}$^,$\aff{University of Florence} \quad Marc Huertas-Company\aff{Instituto de Astrof\`isica de Canarias (IAC)}$^,$\aff{Departamento Astrof\`isica, Universidad de la Laguna}$^,$\aff{Observatoire de Paris, LERMA, PSL University}$^,$\aff{Universit\'e Paris-Cit\'e}\\[4pt]
Maja Jab\l{}o\'nska\aff{UniverseTBD}$^,$\aff{ANU RSAA} \quad Sandor Kruk\aff{European Space Agency} \quad Matthieu Le Lain\aff{IRISA}$^,$\aff{Université Bretagne Sud} \\[4pt]
Sergio José Rodríguez Méndez\aff{UniverseTBD}$^,$\aff{ANU School of Computing} \quad Dimitrios Tanoglidis\aff{UniverseTBD}\\[8pt]
\normalfont $^*$\texttt{shzam@sdf.org} \quad $^\dagger$\texttt{mike@mjjsmith.com}\\[8pt]
\normalfont $^1$UniverseTBD \quad $^2$Indian Institute of Technology Delhi \quad $^3$Intelligent Internet Inc. \\[4pt] 
\normalfont $^4$University of Florence \quad $^5$Instituto de Astrof\'isica de Canarias (IAC)\\[4pt]
\normalfont $^6$Departamento Astrof\'isica, Universidad de la Laguna \\[4pt] \normalfont $^7$Observatoire de Paris, LERMA, PSL University \quad $^8$Universit\'e Paris-Cit\'e \quad $^9$ANU RSAA \\[4pt] 
\normalfont $^{10}$European Space Agency \quad $^{11}$IRISA \quad $^{12}$Université Bretagne Sud\\[4pt]
\normalfont $^{13}$ANU School of Computing
\vspace{-1em}
\end{tabular}}

\iclrfinalcopy 
\begin{document}

\maketitle

\begin{abstract}
We present AstroLLaVA, a vision language model for astronomy that enables interaction with astronomical imagery through natural dialogue. 
By fine-tuning the LLaVA model on a diverse dataset of $\sim$30k images with captions and question-answer pairs sourced from NASA's `Astronomy Picture of the Day', the European Southern Observatory, and the NASA/ESA Hubble Space Telescope, we create a model capable of answering open-ended questions about astronomical concepts depicted visually. 
Our two-stage fine-tuning process adapts the model to both image captioning and visual question answering in the astronomy domain. 
We demonstrate AstroLLaVA's performance on an astronomical visual question answering benchmark and release the model weights, code, and training set to encourage further open source work in this space.
Finally, we suggest a roadmap towards general astronomical data alignment with pre-trained language models, and provide an open space for collaboration towards this end for interested researchers.
\end{abstract}

\section{Language modeling in astronomy}
\label{sec_intro}

Recent advancements in large language models (LLMs) have revolutionised natural language processing, enabling systems to engage in human-like dialogue, answer questions, and assist with tasks across various domains. 
Vision language models (VLMs) extend this capability by grounding the language understanding in visual content, allowing for multimodal interactions. 
These models have gained traction across fields ranging from medical imaging to robotic navigation \citep[e.g.][]{ref_gao2023,ref_van2024}.

In the field of astronomy, domain-specific LLMs like AstroLLaMA have recently emerged.
These models adapt broad language knowledge to the specialised concepts and vocabulary of astrophysics and cosmology \citep{ref_nguyen2023}. 
AstroLLaMA was developed by fine-tuning the LLaMA-7B model \citep{ref_touvron2023} on a curated corpus of astronomy papers, textbooks, and educational web content. 
Subsequent work enhanced AstroLLaMA's conversational ability by further fine-tuning on astronomy question-answering data, enabling the model to provide knowledgeable responses to user queries \citep{ref_perkowski2024}.
Building on this foundation, we present AstroLLaVA, a visual language model tailored for the astronomy domain. 
AstroLLaVA adapts the LLaVA architecture, which combines a vision encoder and language model to enable natural conversations grounded in images \citep{ref_liu2023llava2,ref_liu2023llava}.
By fine-tuning LLaVA on a diverse dataset of public and outreach astronomical images paired with captions and question-answer data, we create a multimodal model that has an inbuilt specialised knowledge of space and astrophysical phenomena.

The key contributions of this work are fourfold: 
a) curating---to our knowledge---the largest ever dataset of outreach-focussed astronomical images paired with high-quality captions and question-answer pairs for fine-tuning; 
b) adapting the LLaVA architecture to the astronomy domain through a two-stage fine-tuning process on image captioning and visual question answering and demonstrating AstroLLaVA's performance on an astronomical visual question answering benchmark; 
c) releasing AstroLLaVA weights and code under the MIT licence to facilitate public and enthusiast engagement with astronomy and neural foundation models;
and d) laying out a roadmap for further research into general astronomical data and text alignment and providing an open space for collaboration between interested researchers and enthusiasts.

\section{Datasets} \label{sec_data}

In this section, we describe in detail the datasets we used to fine-tune AstroLLaVA, with a summary provided in Tab.~\ref{tab_data}. 
We first describe our web-crawled astronomical image/textual caption pairs from NASA's Astronomy Picture of the Day\footnote{\url{https://apod.nasa.gov/apod/}}, then the European Southern Observatory's (ESO) public image archive\footnote{\url{https://www.eso.org/public/images/}}, and then the public ESA/Hubble archive of outreach images from the NASA/ESA Hubble Space Telescope (HST)\footnote{\url{https://esahubble.org/images/}}.
Finally, we explain how we used the textual captions to generate a synthetic conversation dataset.

\begin{wraptable}{l}{0.6\textwidth}
\vspace{-2em}
\caption{
    Summary of our selected datasets. `APOD' is the Astronomy Picture of the Day, `ESO' is the European Southern Observatory, and `HST' is the Hubble Space Telescope.
}
\label{tab_data}
\vskip 0.15in
\begin{center}
\begin{sc}
\begin{tabular}{lcr}
\toprule
Name & Original source & $N$ samples \\
\midrule
APOD        & Official API & 9\,962\\
ESO         & Web scrape & 14\,617\\
HST         & Web scrape & 5\,204\\
\midrule
Total         & -    & 29\,783\\
\bottomrule
\end{tabular}
\end{sc}
\end{center}
\vskip -0.1in
\end{wraptable}

\paragraph{Astronomy Picture of the Day.} \label{sec_apod}

The Astronomy Picture of the Day (APOD) archive boasts the largest collection of annotated public and outreach astronomical images on the Internet.
Since June the 16th, 1995 APOD has posted an astronomy-related image on a near-daily basis, and as of 2025 there are more than 10\,000 image-caption pairs in the NASA APOD archive.

To compile the APOD portion of our dataset we use the API provided by NASA APOD (\url{github.com/nasa/apod-api/}).
We filter the raw dataset to remove all images that are not in \texttt{*.jpeg}, \texttt{*.png}, or \texttt{*.gif} format. 
This filtering leaves us with a total of 9\,962 image-caption pairs.

\paragraph{The ESO and HST archives.} \label{sec_eso}

The ESO image archive contains imagery of astronomical instruments, objects (including but not limited to galaxies, stars, supernovae, and black holes), as well as artistic and computer generated representations of astronomically significant events and objects.
Each entry in this archive is accompanied by a human authored caption explaining the image.

Similarly, ESA/Hubble image archive contains astronomical objects as observed by HST, alongside artistic renditions, imagery of HST itself, promotional imagery of the James Webb Space Telescope, and photography of HST servicing missions.
Every entry in the HST image archive is accompanied with a human authored caption.

We use the \texttt{scrapy} Python package to scrape the ESO and HST archives direct from the websites, resulting in 14\,617 
ESO image-caption pairs and 5\,204 
HST image-caption pairs.

\paragraph{Synthetic conversation generation.} \label{sec_convo}

We use unimodal text GPT-4 to generate synthetic conversations between an `AstroLLaVA' oracle simulacrum and a human inquisitor simulacrum \citep{ref_openai2023gpt4}.
To this end, we prompt \mbox{GPT-4} with an image caption taken from our datasets described above, and ask it to generate between three and five question and answer pairs in a conversational tone.
As we used \mbox{GPT-4} to generate our conversations and not a pre-trained VLM, we do not include our scraped images in our prompting routine, and simply use the text captions as an approximation of our imagery. 
The full prompt used to generate our conversations is shown verbatim in our Github repository.

\section{Methods} \label{sec_model}

\begin{figure}[ptb]
    \centering
    \includegraphics[width=0.8\textwidth]{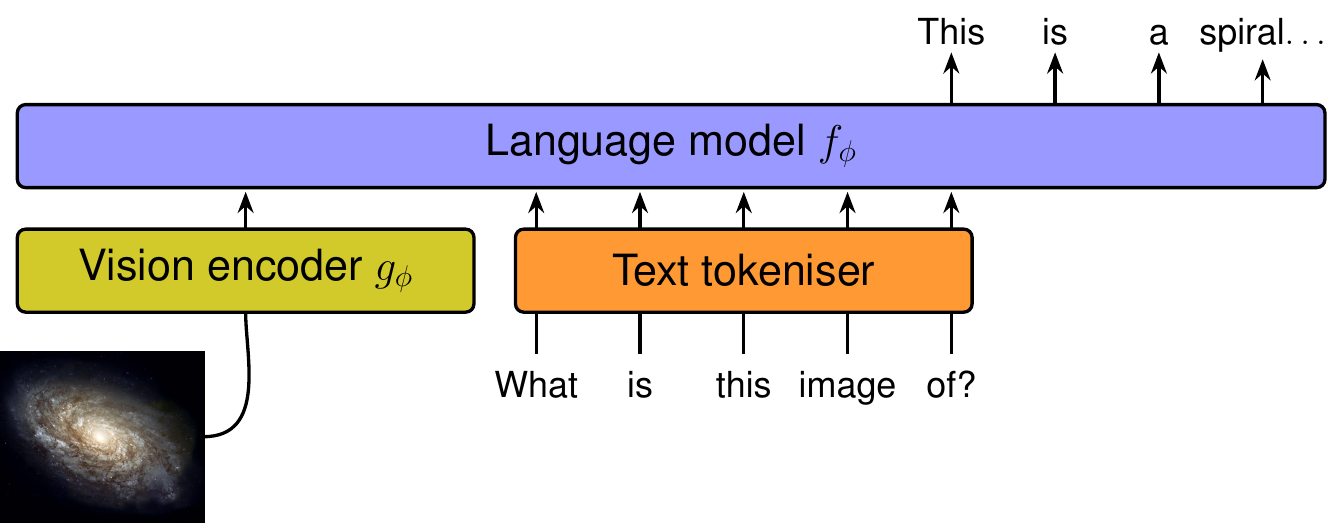}
    \caption{Diagram of our AstroLLaVA model at inference time. $f_\phi$ in this study's case is LLaMA~7B
    \citep{ref_touvron2023}, $g_\phi$ is CLIP ViT-L/14 \citep{ref_radford2021}.}
    \label{fig_llava}
\end{figure}

Our model is based on the general LLaVA 1.5 architecture \citep{ref_liu2023llava2}, while adapting and extending the base LLaVA model in several key ways to imbue astronomical knowledge.
As shown in Fig.~\ref{fig_llava} the core architecture combines a pre-trained vision encoder ($g_\phi$) that processes our astronomical images, a large language model ($f_\phi$) that handles text generation, and learnable projections that bridge the visual and language domains. 
At inference time, the vision encoder first processes the input astronomical image into a set of visual embeddings. 
These embeddings are then projected into the language model's embedding space and prepended to the tokenised text prompt before being passed through the language model.

For the vision encoder $g_\phi$, we use a CLIP ViT-L/14 model \citep{ref_radford2021}, pre-trained on images at a resolution of 336$\times$336 pixels \citep{ref_liu2023llava2}.
The language model $f_\phi$ is based on LLaMA~7B \citep{ref_touvron2023}, which we choose for its strong performance while remaining within reasonable computational constraints. 
To effectively handle astronomical concepts, we augment the standard LLaVA training process with our specialised astronomical dataset described in \S\ref{sec_data}.

Building upon the LLaVA 1.5 pre-training, our model is fine-tuned in two stages:
\begin{enumerate}
    \item We first train only the visual-language projection layers while keeping the pre-trained vision encoder and language model frozen. This stage uses our astronomical image-caption pairs to teach the model to ground astronomical concepts in visual features.
    \item Then we perform instruction tuning using our synthetically generated astronomical QA pairs, fine-tuning the entire model end-to-end.
\end{enumerate}

This two-stage process helps preserve the strong general capabilities of the pre-trained components while adapting them specifically to astronomical data and dialogue. 
In the next section, we compare our model's astronomical knowledge to that of a baseline LLaVA model.

\section{Evaluation}

For our astronomy evaluation baseline we use the Galaxy 10 DECaLS dataset \citep[G10;][]{ref_leung2019,ref_walmsley2022}.
G10 is a MNIST-like dataset containing galaxy images and their corresponding morphological classes (see Fig.~\ref{fig_g10} for class exemplars).
Each model is tasked with generating a brief description of the galaxies in the G10 test set (1770 total images).
To do this we prompt the models with a G10 image and the accompanying text: `Describe the following galaxy image in detail. What type of galaxy is it and what are its key features?'
We then take the cosine distance between embeddings as calculated by \texttt{all-MiniLM-L6-v2}\footnote{\url{https://hf.co/sentence-transformers/all-MiniLM-L6-v2}} for all generated descriptions and the G10 labels (plus chosen descriptive words, see Tab.~\ref{tab_g10}) as our metric.
We compare the 7B and 13B parameter versions of LLaVA 1.5.
We state our results in Tab~\ref{tab_results}.

\begin{wraptable}[12]{r}{0.5\textwidth}
\vspace{-1.5em}
\caption{
    Results of our evaluation on the Galaxy 10 DECaLS dataset.
    `Sim.' is the cosine similarity between embeddings of the generated descriptions and G10 classes plus descriptive words.
}
\label{tab_results}
\begin{center}
\begin{sc}
\begin{tabular}{lr}
\toprule
Model & Sim. ($\uparrow$)\\
\midrule
LLaVA 1.5 7B & 0.594\\
LLaVA 1.5 13B & 0.591\\
\textbf{AstroLLaVA 7B (ours)} & \textbf{0.597}\\
\bottomrule
\end{tabular}
\end{sc}
\end{center}
\end{wraptable}
The similar performance between LLaVA 1.5 7B, LLaVA 1.5 13B, and AstroLLaVA 7B on the G10 dataset classification task could stem from limitations in the underlying CLIP vision encoder that all three models share, or from the outreach-heavy dataset that we have used to fine-tune AstroLLaVA in this work.
Firstly, since all these models use the same vision encoder to extract features from the galaxy images, they are fundamentally limited by the information captured by these CLIP embeddings:
this is the case even with AstroLLaVA's specialised astronomical training and LLaVA-13B's additional parameters.
This suggests that the performance bottleneck could be the visual encoder rather than language modelling capacity or domain adaptation.
Secondly, the compiled dataset consisting of APOD, ESO, and HST imagery may be too focussed on outreach to provide a strong fine-tuning signal for the galaxy classification task.
Therefore, future improvements will concentrate on training improved visual encoders specifically optimised for scientific astronomical imagery (and further modalities, see the next section) rather than simply fine-tuning the projection layers and language model components.

\section{Conclusions and future work}

AstroLLaVA lights a viable path towards an exciting destination: extending this model beyond imagery and text to create a truly multi-modal astronomical foundation model. 
Just as PAPERCLIP \citep{ref_mishra2024} demonstrates the power of leveraging aligned proposal text to ground visual representations, we envision applying similar techniques to align a broader range of astronomical data types within a shared embedding space. 
Astronomical objects naturally manifest across multiple observational modes---from photometric time series and spectra to radio interferometry and N-dimensional data cubes \citep{ref_MMU2024}.
A model that can seamlessly reason across these modalities would be transformative for modern survey astronomy.

Beyond its role in expanding multi-modal astronomical data analysis, models like AstroLLaVA can significantly enhance data accessibility through multi-sensory perception. By integrating machine-generated voice descriptions with data-sonification frameworks \citep[e.g.][]{ref_ginolfi2024}, astronomical imagery can become more accessible to blind and visually impaired individuals. This combination allows users to hear and interact with data through both structured descriptions and auditory representations, fostering a richer and more inclusive experience. More broadly, multi-sensory approaches can enhance engagement with astronomical data for all users, offering novel ways to explore complex astrophysical information beyond traditional visual analysis.

AstroLLaVA has so far been trained primarily on public-facing astronomical imagery and captions from outreach-oriented sources such as APOD, ESO, and HST. Future work will extend this foundation by incorporating scientific-grade datasets from astronomical literature and archives. Fine-tuning on such domain-specific data would enable more advanced tasks like galaxy classification, anomaly detection, and astronomical object detection, better supporting researchers in scientific workflows \citep{ref_tanoglidis2024,ref_fu2024,ref_riggi2025}.

To realise these visions, we propose extending the AstroLLaVA architecture to handle arbitrary astronomical data tensors through modality-specific encoders that project into a common latent space bridged by language. 
These encoders could handle---for instance---SDSS spectra \citep{ref_blanton2017}, TESS light curves (time series) \citep{ref_ricker2014}, MaNGA data cubes \citep{ref_bundy2014}, and other key observational products.
The language model would then serve as a \emph{lingua franca} interface for querying and reasoning about patterns across all modalities simultaneously.
Progress toward this ambitious goal will require substantial community involvement \citep{ref_phang2022}.
To this end, we are not only releasing our code and models, but also welcome interested parties to an open collaboration space on the UniverseTBD Discord server to further co-ordinate development efforts.
We envision this as a truly community-driven effort to build the next generation of multi-modal astronomical foundation models.

\subsubsection*{Carbon emissions}

The training of deep learning models requires considerable energy, which can contribute to carbon emissions. 
To counteract further emission from unnecessary retraining we follow the recommendations of \citet{ref_strubell2019} and make available our AstroLLaVA model weights and inference code. 
While this study made use of the 100\% renewable energy-powered ITER Teide HPC cluster \citep{ref_mampaso2022} for training, for the sake of transparency we estimate and show our energy usage during the training of AstroLLaVA here:
AstroLLaVA takes 5 hours to train on 4xA100-40G NVIDIA GPUs, which corresponds to 5\,kWh of energy use according to the Machine Learning CO$_2$ Impact Calculator (\url{https://mlco2.github.io/impact}).

\subsubsection*{Open artefacts}

Our training dataset is available at:\\
\url{https://w3id.org/UniverseTBD/AstroLLaVA/dataset}\\
Our code is available at:\\
\url{https://w3id.org/UniverseTBD/AstroLLaVA}\\
We welcome collaboration via:\\
\url{https://discord.gg/PUR2FbFRZ4}

\subsubsection*{Acknowledgments}
We would like to thank the Microsoft Accelerating Foundation Models Research program for providing cloud compute credits for generating our synthetic conversations.
We thank the \href{https://teidehpc.iter.es/en/home/}{ITER Teide HPC cluster} for their support. 

\bibliography{main}
\bibliographystyle{iclr2025_conference}

\appendix

\vfill
\pagebreak

\section{Full prompt} \label{sec_prompt}

\begin{figure*}[h!]
\begin{lstlisting}
PROMPT_CONV = """
You are AstroLLaVA, an AI assistant with a special knowledge of astronomical topics. 
You are provided with the following description of a NASA "Astronomical Picture of the Day" image.
Unfortunately the original image is not available

------
What does a black hole look like? To find out, radio telescopes from around the Earth coordinated observations of black holes with the largest known event horizons on the sky. Alone, black holes are just black, but these monster attractors are known to be surrounded by glowing gas. This first image resolves the area around the black hole at the center of galaxy M87 on a scale below that expected for its event horizon. Pictured, the dark central region is not the event horizon, but rather the black hole's shadow -- the central region of emitting gas darkened by the central black hole's gravity. The size and shape of the shadow is determined by bright gas near the event horizon, by strong gravitational lensing deflections, and by the black hole's spin. In resolving this black hole's shadow, the Event Horizon Telescope (EHT) bolstered evidence that Einstein's gravity works even in extreme regions, and gave clear evidence that M87 has a central spinning black hole of about 6 billion solar masses. Since releasing this featured image in 2019, the EHT has expanded to include more telescopes, observe more black holes, track polarized light,and is working to observe the immediately vicinity of the black hole in the center of our Milky Way Galaxy.
------
Design a conversation between you and a person asking about the above photo. 
The answers should be in a tone that a visual AI assistant is seeing the image and answering the question. Ask diverse questions and give corresponding answers.

Below are the requirements for generating the questions:
1. Do not ask the redshift of galaxies, as you can not measure the redshift of the galaxy just from the image.
2. Focus on the visual aspects of the image such as size, position of the object in the image, special features, morphology, artifacts that can be inferred without the text information.
3. Make the questions casual and conversational.
4. Do not attempt to name objects in the image. Please use a generic term like "galaxy", "planet", or "star".

Below are the requirements for generating the answers:
1. Avoid quoting or referring to specific facts, terms, abbreviations, dates, numbers, or names.
2. Do not use phrases like "mentioned", "caption", "context", "description", "text" in the conversation. Instead, refer to the information as being "in the image" or "in the picture".
3. Avoid referring to "based on the conversation" in answers.
4. Avoid giving answers when there is not enough information.
5. DO NOT use phrases like "mentioned", "caption", "context", "description", "text" in the conversation. YOU MUST refer to the information as being "in the image."
4. DO NOT attempt to name objects in the image. Please use a generic term like "galaxy", "planet", or "star".

Please come up with a set of between three and six user questions and assistant answers about that image. 

Here is one example of questions and answers when the image is of the Fornax Cluster:

human: Hey astrollava, what am I looking at here?
astrollava: Hello human! You're looking at an image of the Fornax Cluster, a nearby cluster of galaxies. This cluster is named after the southern constellation Fornax, where most of its galaxies are found. The yellowish splotches are predominantly elliptical galaxies in the cluster.
human: Those bright dots scattered across the image, are they also galaxies?
astrollava: Good observation! While most of the yellowish splotches are galaxies, the brighter and sharper dots you're seeing are actually spiky foreground stars. They are much closer to us than the galaxies in the background.
human: How far is this cluster from us?
astrollava: The Fornax Cluster is approximately 62 million light-years away from us. To give you some perspective, it's almost 20 times more distant than our neighboring Andromeda Galaxy.
human: Thank you, astrollava. Your explanations have been truly enlightening!
astrollava: You're welcome, human! I'm always here to help you navigate the wonders of the cosmos. If you have any more questions, feel free to ask!

If there is not enough information in the conversation, respond with something along the lines of 'I don't have enough information to answer that question.'.

Please respond with a json file format like this:

[
    {
        "from": "human",
        "value": human question,
    },
    {
        "from": "astrollava",
        "value": astrollava response,
    }
    
]

where the human questions and astrollava responses are the question and answer generated.
Please make the human questions and astrollava answers casual and conversational.
"""
\end{lstlisting}
\caption{Example prompt to generate a conversation for the APOD published on 2022-05-01, titled `First Horizon-Scale Image of a Black Hole'. Similar prompts are used for all our datasets.}
\label{fig_prompt}
\end{figure*}

\section{Galaxy 10 DECaLS example imagery}

\begin{figure}[h]
    \includegraphics[width=\textwidth]{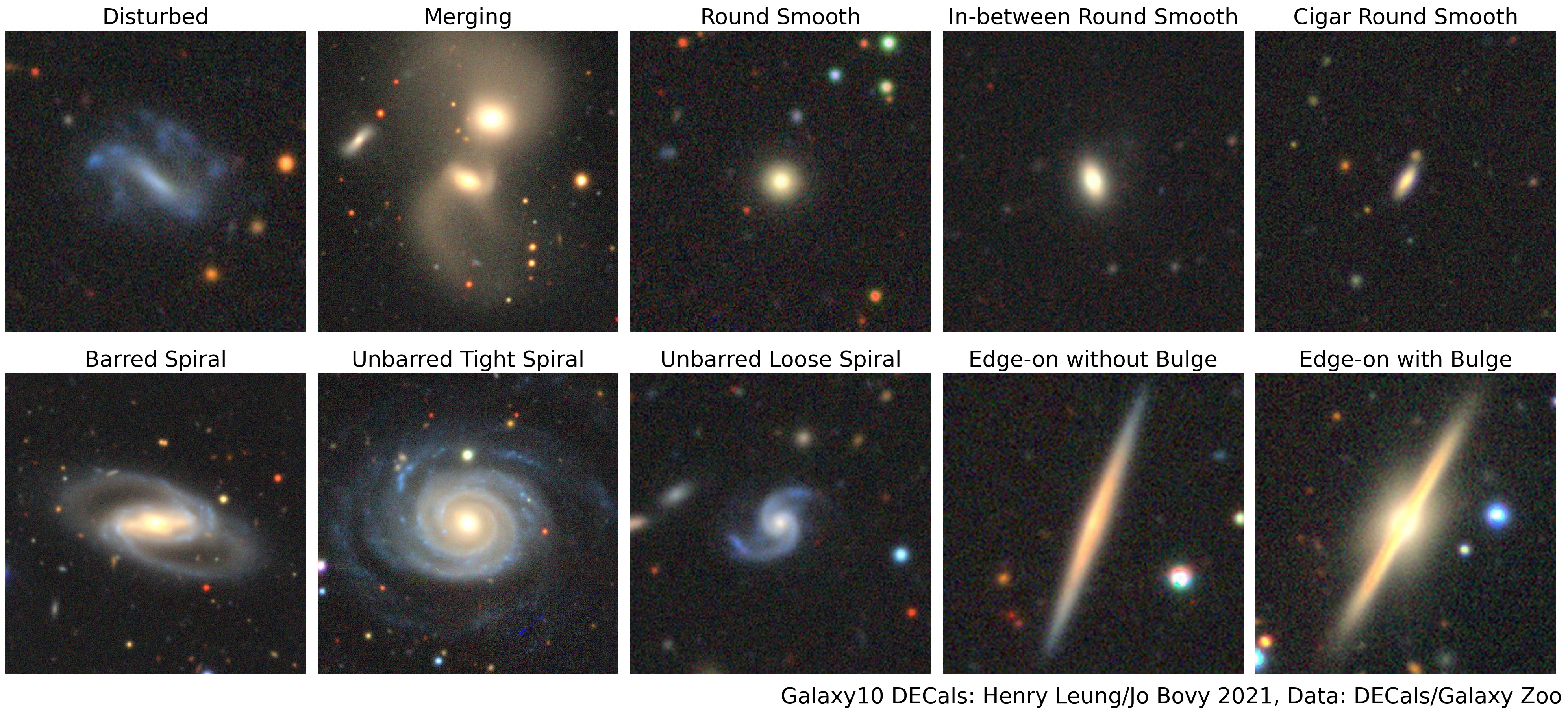}
    \caption{Exemplar galaxy images for each class in the `Galaxy 10 DECaLS' dataset. This figure is taken from \url{https://github.com/henrysky/Galaxy10}.}
    \label{fig_g10}
\end{figure}

\begin{table}[h]
\caption{Classification of galaxies in Galaxy 10 DECaLS and the selected distinguishing features used in our semantic similarity evaluations.}
\begin{center}
\begin{sc}
\begin{tabular}{lp{0.48\textwidth}}
\toprule
Class & Features \\
\midrule
Disturbed Galaxies & irregular shape, asymmetric, disturbed structure, unusual patterns, deformed \\
Merging Galaxies & multiple cores, interaction, tidal tails, bridges, overlapping \\
Round Smooth Galaxies & circular, elliptical, smooth, regular \\
In-between Round Smooth Galaxies & somewhat round, partially smooth, mild irregularity, intermediate shape \\
Cigar Shaped Smooth Galaxies & elongated, cigar-shaped, smooth, uniform, linear structure \\
Barred Spiral Galaxies & bar structure, spiral arms, central bar \\
Unbarred Tight Spiral Galaxies & tight spiral arms, no bar, well-defined arms, compact structure \\
Unbarred Loose Spiral Galaxies & loose spiral arms, no bar, open structure, widely spaced arms \\
Edge-on Galaxies without Bulge & edge-on, thin disk, no central bulge, linear structure \\
Edge-on Galaxies with Bulge & edge-on, central bulge, thick centre, disk structure \\
\bottomrule
\end{tabular}
\end{sc}
\end{center}
\label{tab_g10}
\end{table}

\end{document}